 \documentclass[final,5p,times,twocolumn,authoryear]{elsarticle}

\usepackage{amssymb}
\usepackage{lipsum}
\usepackage{xcolor}
\usepackage{braket}
\usepackage{amsmath}

\journal{Journal of Molecular Liquids}

\begin{document}

\begin{frontmatter}



\title{Thermophysical Properties of Molten FLiNaK: a Moment Tensor Potential Approach}


\author[first]{Nikita Rybin}
\affiliation[first]{organization={Skolkovo Institute of Science and Technology},
            addressline={Bolshoy Boulevard 30, bld. 1}, 
            city={Moscow},
            postcode={121205}, 
            country={Russia}}
            
\author[second]{Dmitrii Maksimov}
\affiliation[second]{organization={Institute of High Temperature Electrochemistry, Ural Branch of the Russian Academy of Sciences},
            addressline={Academicheskaya Str. 20}, 
            city={Yekaterinburg},
            postcode={620066}, 
            country={Russia}}

\author[second]{Yuriy Zaikov}

\author[first]{Alexander Shapeev}

\begin{abstract}
Fluoride salts demonstrate significant potential for applications in next-generation nuclear reactors, necessitating a comprehensive understanding of their thermophysical properties for technological advancements. Experimental measurement of these properties poses challenges, due to factors such as high temperatures, impurity control, and corrosion. Consequently, precise computational modeling methods become essential for predicting the thermophysical properties of molten salts. In this work, we performed molecular dynamics (MD) simulations of several thermophysical properties of the eutectic salt mixture LiF-NaF-KF (FLiNaK) melt, including density, self-diffusion coefficients, viscosity, and thermal conductivity. We demonstrated the successful application of moment tensor potentials (MTP) as an accurate model for interatomic interactions in FLiNaK. Our results on thermophysical properties calculations exhibit strong agreement with experimental data. An important aspect of our methodology is the incorporation of an active learning scheme, which enables the generation of a robust and accurate potential, while maintaining a moderate-sized training dataset.
\end{abstract}




\end{frontmatter}




\section{Introduction}
\label{introduction}

Molten salts play a pivotal role in various industrial applications, including  molten salt batteries~\cite{Ong2020, BELL2019109328, Kaixuan2022, Parasotchenko2023ChoiceOT, Leblanck2010}, molten salt reactor systems~\cite{BENES200922, MAGNUSSON2020107608}, as well as in pyroprocessing methods for recycling nuclear fuel~\cite{ZHITKOV2020123, ZHITKOV2022653, MULLABAEV2022965}. Most important characteristics of melts from an application standpoint are temperature-dependent values of density, diffusivity, viscosity and thermal conductivity, which are usually measured experimentally. Despite the availability of numerous experimental measurements for different molten salts, reported thermophysical properties exhibit significant scatter and uncertainties, reaching up to 20\%~\cite{Romatoski2017}. This variability is attributed to challenges associated with impurities, high-temperature measurements, and deviations in composition.

As an alternative to experimental methods, accurate theoretical approaches based on \textit{ab initio} molecular dynamics (AIMD) simulations have emerged as valuable tools for evaluating thermophysical properties in well-controlled conditions~\cite{Porter2022}. While AIMD has significantly advanced our understanding of the structure of various molten salts, essential properties such as diffusion coefficients, viscosity, and thermal conductivity necessitate larger unit cell sizes and extended simulation times for relevant statistical analysis \cite{Gheribi2014, Jabbari2017}. Thus, there is a critical need for an efficient computational method to reliably predict the thermophysical properties of molten salts.

An alternative computational approach involves MD simulations, where interatomic interaction potential is fitted to first-principles calculations or experimental data~\cite{Salanne2009, doi:10.1021/acs.jpcb.2c06915, MAXWELL2022153633}. This overcomes limitations related to simulation cell sizes and time scales, enabling simulations of larger structures for more extended periods compared to AIMD. Properties derived from classical MD simulations generally achieve sufficient numerical precision, with errors dominated by the accuracy of the underlying interatomic potentials~\cite{Lu2021, Lee2021}.

In the last decade, significant progress has been made in developing the so-called machine learning interatomic potentials (MLIPs)~\cite{https://doi.org/10.1002/adma.201902765}.
MLIPs have demonstrated promise for MD simulations of molten salts modelling with near \textit{ab initio} accuracy, on time and length scales comparable to traditional interatomic potentials~\cite{Liang2020, Feng2022, Lam2021, Li2021, Rodriguez2021, Attarian2022}. The moment tensor potential~\cite{Shapeev2016} (MTP) was shown to be among the most efficient (in terms of data utilization) and accurate machine-learned models for interatomic interactions~\cite{Zuo2020}. In the case of molten LiF-BeF$_{2}$ mixture, it was recently shown that MTP can be used to compute thermophysical properties with both high precision and low data utilization~\cite{Attarian2022}.

In this study, we employ MTPs to calculate thermophysical properties of molten FLiNaK at eutectic composition (46.5-11.5-42 mol\%). FLiNaK is among the most promising and well-investigated salts~\cite{Locatelli2013, ma16114197, LIZIN2017375}, with a lot of experimental and theoretical data. We fitted an MTP to approximate the potential energy surface of FLiNaK on the data calculated from first principles in the dispersion-corrected Density-Functional Theory (DFT-D3) framework. As will be demonstrated, MTP model used in couple with MD simulations allows to obtain thermophysical properties of FLiNaK in a good agreement with reported literature data, both experimental and theoretical. 

\section{Methodology}

\subsection{Moment Tensor Potential}

In this work, we used MTP approach implemented in the MLIP-2 package~\cite{Novikov2021} to investigate the thermophysical properties of molten FLiNaK at finite temperatures. The potential energy of an atomic system as described by the MTP interatomic potential is defined as a sum of the energies of atomic environments of the individual atoms:

\[
    E_{\text{MTP}} = \sum_{i=1}^{N} V(n_{i}),
\]
where the index $i$ label $N$ atoms of the system, and $n_{i}$ describes the local atomic neighborhood around atom \textit{i} within a certain cutoff radius $R_\text{cut}$ and the function $V$ is the moment tensor potential: 
\[
    V(n_{i}) = \sum_{\alpha} \xi_{\alpha} B_{\alpha}(n_{i}),
\]
where $\xi_{\alpha}$ are the fitting parameters and $B_{\alpha}(n_{i})$ are the basis functions that will be defined below. Moment tensors descriptors are used as representations of atomic environments and defined as: 
\[
    M_{\mu, \nu}\left({n}_i\right)=\sum_j f_\mu\left(\left|r_{i j}\right|, z_i, z_j\right) \underbrace{r_{i j} \otimes \ldots \otimes r_{i j}}_{\nu \text { times }},
\]
where the index $j$ goes through all the neighbors of atom $i$. The symbol ``$\otimes$'' stands for the outer product of vectors, thus ${r}_{i j} \otimes \cdots \otimes {r}_{i j}$ is the tensor of rank $\nu$ encoding the angular part which itself resembles moments of inertia. 
The function $f_\mu$ represents the radial component of the moment tensor:
\[
f_\mu\left(\left|r_{i j}\right|, z_i, z_j\right)=\sum_k c_{\mu, z_i, z_j}^{(k)} Q^{(k)}(r),
\]
where $z_i$ and $z_j$ denote the atomic species of atoms $i$ and $j$, respectively, $r_{ij}$ describes the positioning of atom $j$ relative to atom $i$, $c_{\mu, z_i, z_j}^{(k)}$ are the fitting parameters and 
\[
Q^{(k)}(r):=T_k(r)\left(R_{\text {cut }}-r\right)^2
\]
are the radial functions consisting of the Chebyshev polynomials $T_k(r)$ on the interval $[R_\text{min},  R_\text{cut}]$ with the term ($R_\text{cut} - r)^2$ that is introduced to ensure a smooth cut-off to zero. The descriptors $M_{\mu, \nu}$ taking $\nu$ equal to $0, 1, 2, \ldots$ are tensors of different ranks that allow to define basis functions as all possible contractions of these tensors to a scalar, for instance:
\[
\begin{aligned}
& B_0\left({n}_i\right)=M_{0,0}\left({n}_i\right), \\
& B_1\left({n}_i\right)=M_{0,1}\left({n}_i\right) \cdot M_{0,1}\left({n}_i\right), \\
& B_2\left({n}_i\right)=M_{0,0}\left({n}_i\right)\left(M_{0,2}\left({n}_i\right): M_{0,2}\left({n}_i\right)\right) .
\end{aligned}
\]
Therefore the level of $M_{\mu, \nu}$ is defined by ${\rm lev}M_{\mu, \nu}$ = 2$\mu$ + $\nu$ and if $B_{\alpha}$ is obtained from $M_{\mu_1, \nu_1}$ , $M_{\mu_2, \nu_2}$ ,
$\dots$, then ${\rm lev}B_{\alpha}$ = $(2\mu_1 + \nu_1)$ + $(2\mu_2 + \nu_2)$ + $\dots$ . By including all basis functions such that ${\rm lev}B_{\alpha} < d$ we obtain the moment tensor potential of level $d$, which we denote as MTP$_d$.

\subsection{Dataset Generation and Potential Fitting}

The MTP utilized in this study is trained using data computed within the Density-Functional Theory (DFT) framework. All DFT calculations were carried out using VASP (Vienna \textit{ab initio} simulations package)~\cite{Kresse1996} with the projector augmented wave method~\cite{Kresse1999}. The Perdew-Burke-Ernzerhof generalized gradient approximation (PBE-GGA)~\cite{Perdew1996} was employed for the exchange–correlation functional, and the DFT-D3 method~\cite{Grimme2010} was utilized to account for dispersion forces.

The initial dataset was generated from four independent AIMD trajectories (at temperatures $T=800$~K, $1000$~K, $1200$~K, and $1400$~K), which were conducted in the isothermal-isobaric ensemble (NPT) using the Nos\'{e}-Hoover thermostat~\cite{Nose1984}. Each AIMD trajectory was simulated for 2~ps with a 1~fs time step and for supercells containing 56 atoms (28~F atoms, 13~Li atoms, 3~Na atoms, and 12~K atoms) to closely mimic eutectic composition. The plane-wave basis set had an energy cutoff of 550~eV, and a single gamma point was used to sample the Brillouin zone. The initial structures for AIMD were generated by randomly distributing atoms in the cubic cell, which had the volume equal to the total volume of spheres, each with a radius matching the van~der~Waals radii of the respective atoms. The first picosecond of each AIMD simulation was discarded, and the second picosecond was subsampled with the 5~fs time intervals, resulting in a set of 800 samples. These samples were employed to train an initial MTP of level 14.

The root-mean-squared-errors (RMSEs) on this dataset are 2.1~meV/atom for energies and 48~meV/$\AA$ for forces. Previous studies~\cite{Feng2022, Lam2021} have indicated that RMSE of energy smaller than 5~meV/atom and RMSEs on forces smaller than 100~meV/$\AA$ are generally sufficient for predicting properties such as density, radial distribution functions, diffusion coefficient, and viscosity of molten salts. However, achieving this accuracy does not by itself ensure the robustness (ability to run long MD) of the potential. Generally, two strategies are employed to train a robust MLIP capable of capturing the entire range of local atomic environments: (i) including a variety of different systems which requires lengthy AIMD simulations, and (ii) utilizing an active learning (active sampling, learning on-the-fly) strategy to selectively add data points on which the MLIP extrapolates significantly, i.e., the prediction of energies and forces of these structures is done with high uncertainty.

\begin{table}[ht]
	\centering
\caption{Potential energy and force RMSEs on training and test sets for the fitted Moment Tensor Potential.} 
\begin{tabular}{cccll}
\cline{1-3}
\multicolumn{1}{c|}{}       & \multicolumn{1}{c|}{Train} & Test                 &  &  \\ \cline{1-3} 
\multicolumn{1}{c|}{Energy (meV/atom)} & \multicolumn{1}{c|}{1.83} & 1.93                 &  &  \\
\multicolumn{1}{c|}{Forces (meV/$\AA$)} & \multicolumn{1}{c|}{39}  & 39                &  &  \\ \cline{1-3}
\multicolumn{1}{l}{}        & \multicolumn{1}{l}{}       & \multicolumn{1}{l}{} &  & 
\end{tabular}
\label{tab:test_train_errors}
\end{table}

\begin{figure*}[h!]
	\centering
	\begin{minipage}[h]{0.50\linewidth}
		\center{\includegraphics[trim={0cm 0cm 0cm 0cm}, clip, width=1\linewidth]{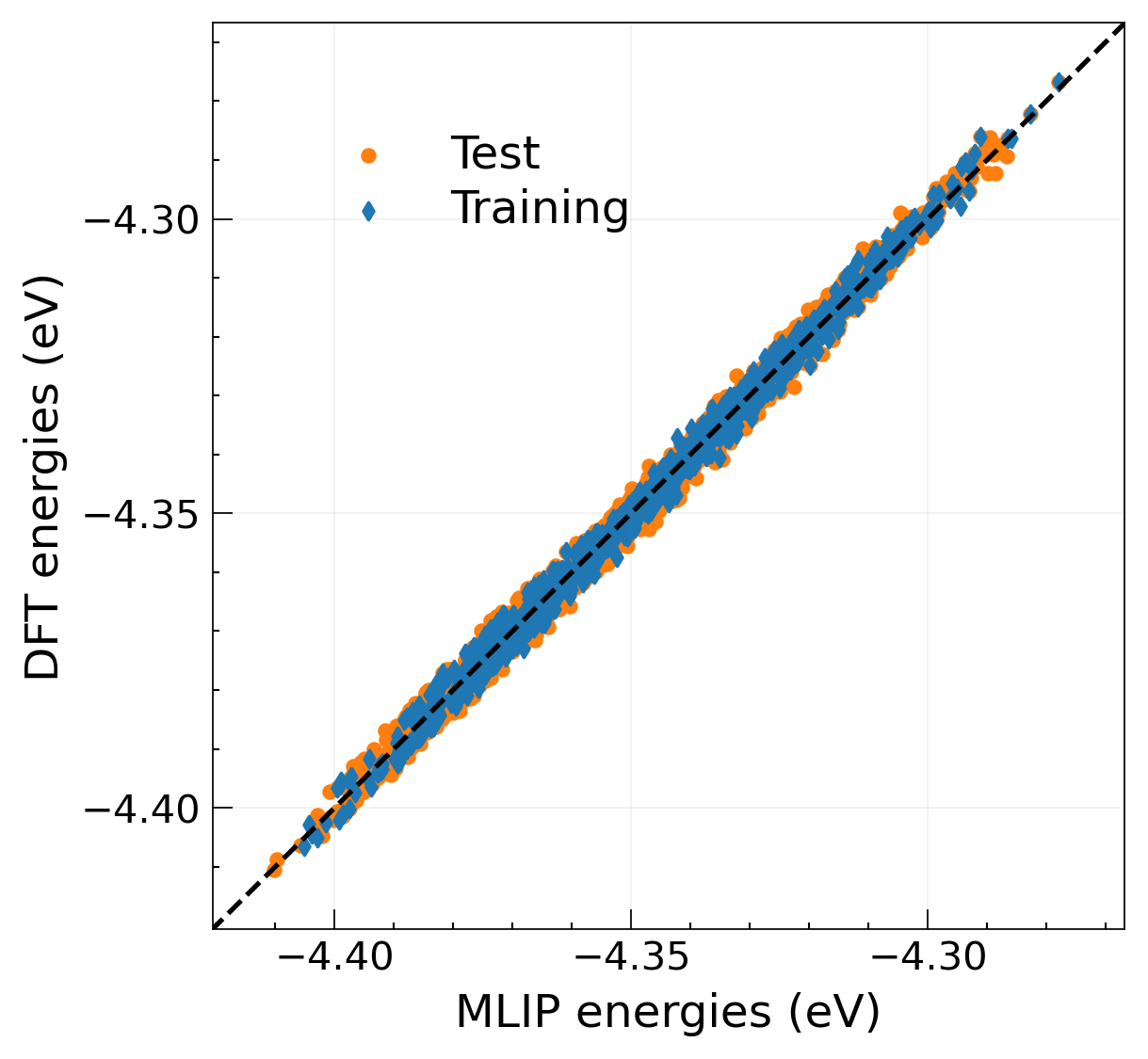}} \\ (a)
	\end{minipage}
 	\begin{minipage}[h]{0.48\linewidth}
		\center{\includegraphics[trim={0cm 0cm 0cm 0cm}, clip, width=1\linewidth]{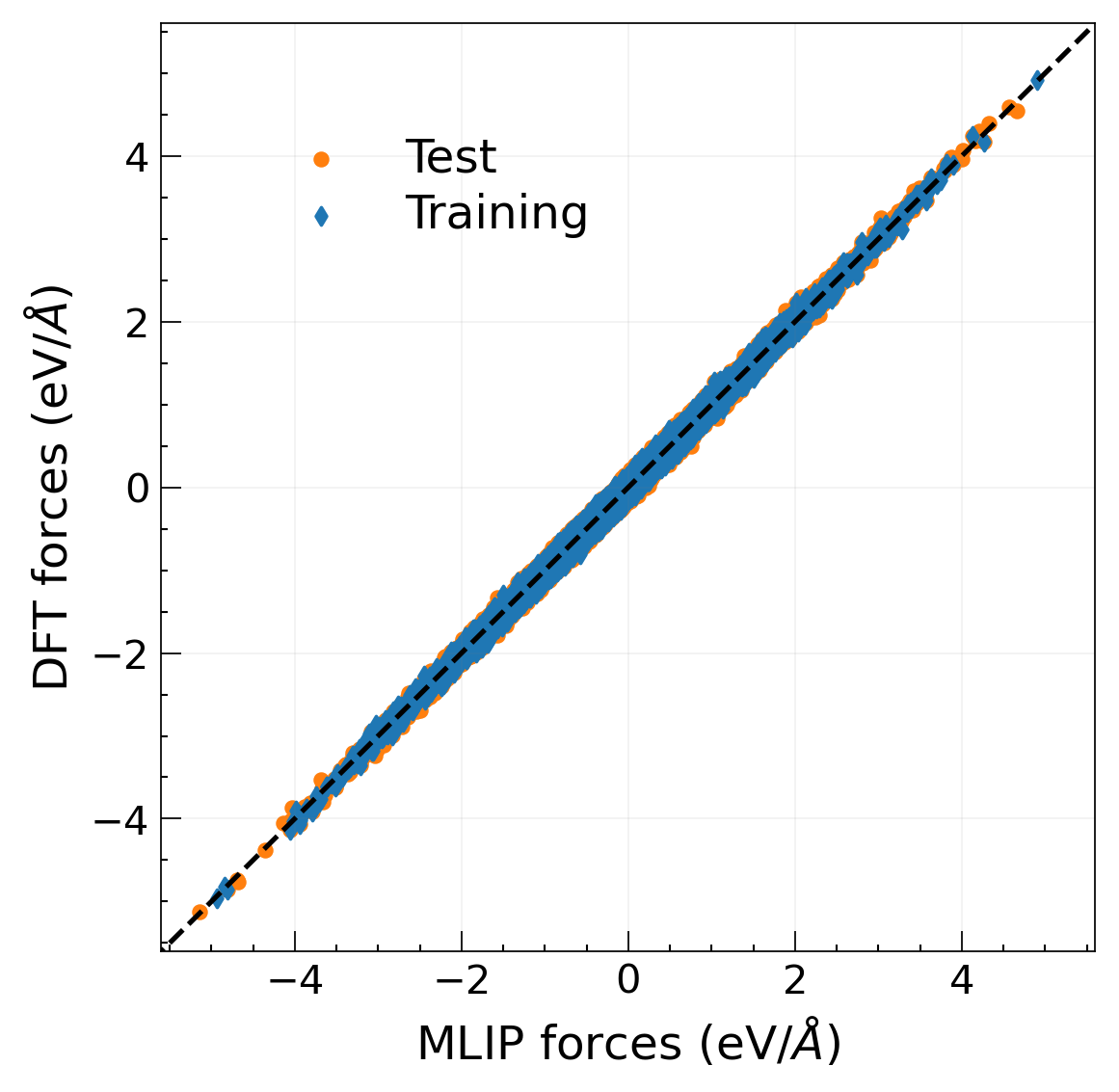}} \\ (b) 
	\end{minipage}
	\caption{Parity plots of the energies (a) and forces (b) for training (blue) and test (orange) sets. The diagonal line in each figure shows the perfect fit.}
	\label{fig:ef_parity_plots}
\end{figure*}

For the purpose of methodological validation, we have explored both strategies. We extended AIMD simulations by an additional 3~ps at each temperature, yielding (after subsampling) a dataset of 3200 samples. We shuffled this dataset and divided it into the training (20\% of the total dataset) and test sets (remaining 80\% of the dataset). Fig.~\ref{fig:ef_parity_plots} presents a comparative analysis of DFT and MTP calculated energies and forces for both training and test sets. As presented in Tab.~\ref{tab:test_train_errors}, the RMSE of MTP on energies is just 1.93 meV/atom, and on forces, it is 39 meV/$\AA$. The displayed RMSE values in Fig.~\ref{fig:ef_parity_plots} demonstrate that the MTP, trained on a 20\% subset of configurations derived from AIMD simulations, accurately predicts the energies and forces of the remaining 80\% (test set) with comparable errors to those of the training set. Low errors on the test set, suggests that the MTP is well-fitted and likely capable of extrapolating to diverse structures and chemical environments. 

However, even with these promising results, an ability to run long MD simulations without failure remains to be demonstrated. To assess stability of the fitted potential, we conducted MD simulations using MTP as the model for interatomic interactions in the micro-canonical (NVE) ensemble within LAMMPS (Large-scale Atomic/Molecular Massively Parallel Simulator)~\cite{Thompson2022}. Unfortunately, these simulations proved to be unstable, exhibiting instability after several tens of picoseconds. Notably, extending the potential training to the entire dataset did not alleviate this issue. This outcome underscores that while low energy and force errors are achieved, they do not guarantee the robust applicability of a MLIP for practical MD simulations, even under the same thermodynamic conditions as those employed in the training data.

We now revisit the active learning strategy, which means interactively selecting a diverse, but minimal, set of training data in the feature space to effectively fit the potential. Various active learning schemes are employed in the development of machine learning (ML) potentials \cite{Zhang2019, Sivaraman2020}. In our study, we utilized the D-optimality-based active learning procedure developed in~\cite{Podryabinkin2017} and available in the MLIP-2 package.

Within the active learning algorithm, we initiate MD calculations in LAMMPS using a pretrained MTP. At each step of the MD simulations, the algorithm assesses the extrapolation grade $\gamma$ of the atomic configuration based solely on atomic coordinates. Configurations with $\gamma > 2$ are added to the preselected set. When $\gamma$ exceeds 10, the MD simulation halts, and all sufficiently different configurations from the preselected set are incorporated into the training, followed by the refitting of the potential. This procedure repeats until MD simulations can run without failure for 200~ps. In our case, MTP achieved robustness with a training set comprising 880 samples, indicating that only a small portion of the actively selected configurations required single-point DFT calculations. The potential trained in this manner demonstrates the ability to robustly conduct MD simulations for at least hundreds of picoseconds, showcasing the effectiveness of the active learning procedure in terms of reducing the necessary DFT data for training and enhancing potential robustness. The training set generated during the active learning procedure can later be used to train a MTP with enhanced accuracy, as will be demonstrated in this study. 

Thus, the active learning scheme plays a crucial role in enhancing the robustness of any MLIP, and MTP, in particular. Its employment not only ensures the applicability of MTP at large length scales and extended time scales in the MTP-MD simulations, but also significantly reduces the number of required DFT calculations. In the subsequent sections of the paper, we present the results of our calculations on the thermophysical properties of FLiNaK. Specifically, we assess the MTP's capability to accurately calculate the temperature dependencies in density, diffusivity, viscosity, and thermal conductivity.

\section{Results}

\subsection{Radial Distribution Functions and Self-Diffusion Coefficients}

\begin{figure}
	\centering 
	\includegraphics[trim={1cm 0cm 18cm 3cm}, clip, width=0.86\linewidth]{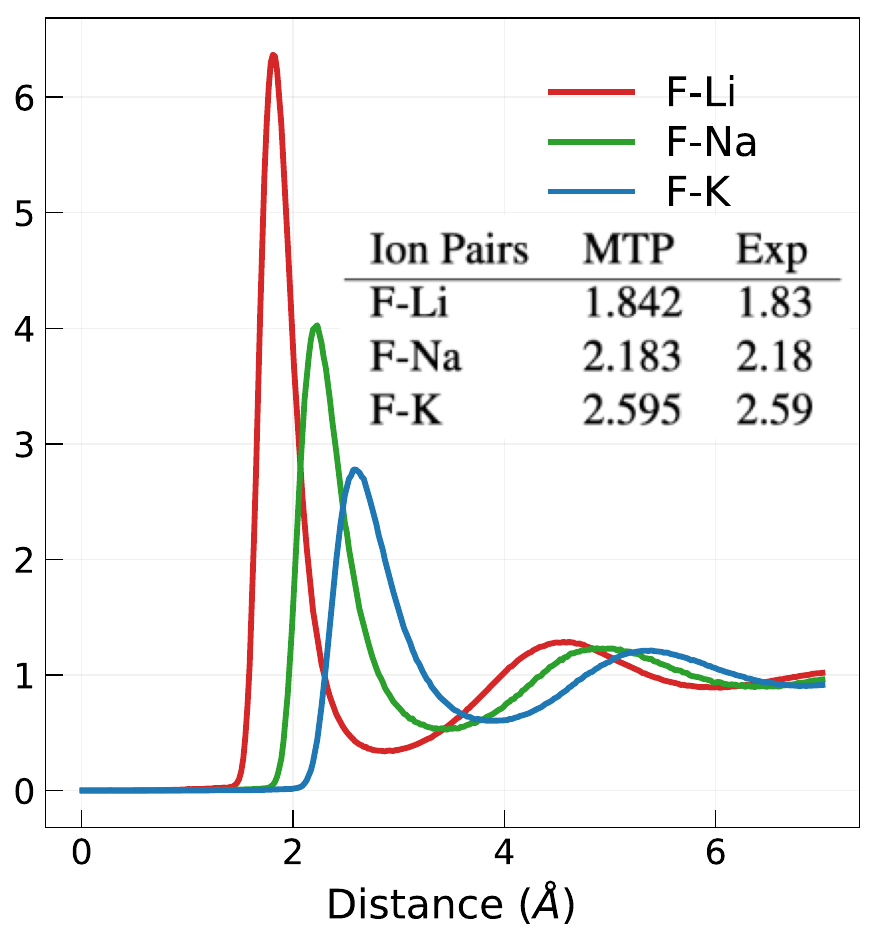}	
	\caption{Radial distribution functions, extracted from MTP-MD at temperature T=793~K for F-Li, F-Na, and F-K ion pairs. Experimental nearest-neighbor peak distance for aforementioned ion pairs was determined from X-ray scattering experiment~\cite{Igarashi1988}.} 
	\label{fig:rdf}
\end{figure}

To assess how well the developed MTP can predict local structural features of molten FLiNaK, we first calculated radial distribution functions (RDF) and diffusion coefficients of ions. The smallest unit cell of molten FLiNaK in an eutectic composition contains 400 atoms: 93~Li atoms, 200~F atoms, 23~Na atoms, and 84~K atoms, respectively. For MTP-MD simulations, the simulation cell was replicated in a 2$\times$2$\times$2 arrangement, which resulted in a supercell with 3200 atoms. The usage of larger supercells yields better statistics of calculations and MTP-MD allows to perform long simulations, which is inaccessible by AIMD. The first RDF peaks for F-Li, F-Na, and F-K are shown in Fig.~\ref{fig:rdf}. RDF for Li-F shows a sharp peak at 1.842~$\AA$ and then rapidly decays. This suggests a strong bonding between Li and F within FLiNaK. On the other hand, the RDF for K-F has a wider first peak at 2.595~$\AA$ that does not decay to zero. This result suggests that the first nearest neighbor shell for K and F is more diffusive and not as well defined as for Li and F, which further indicated that the bonding between K and F in FLiNaK is not as strong as between Li and F. The values of the calculated first peak distances obtained using MTP-MD for F-Li, F-Na, and F-K are close to the experimental values~\cite{Igarashi1988} as shown in an inset of Fig.~\ref{fig:rdf}. This indicates that MTP can accurately capture the local environment of ions in molten FLiNaK.

\begin{figure*}[h!]
	\centering
	\begin{minipage}[h]{0.49\linewidth}
		\center{\includegraphics[trim={0cm 0cm 0cm 0cm}, clip, width=0.95\linewidth]{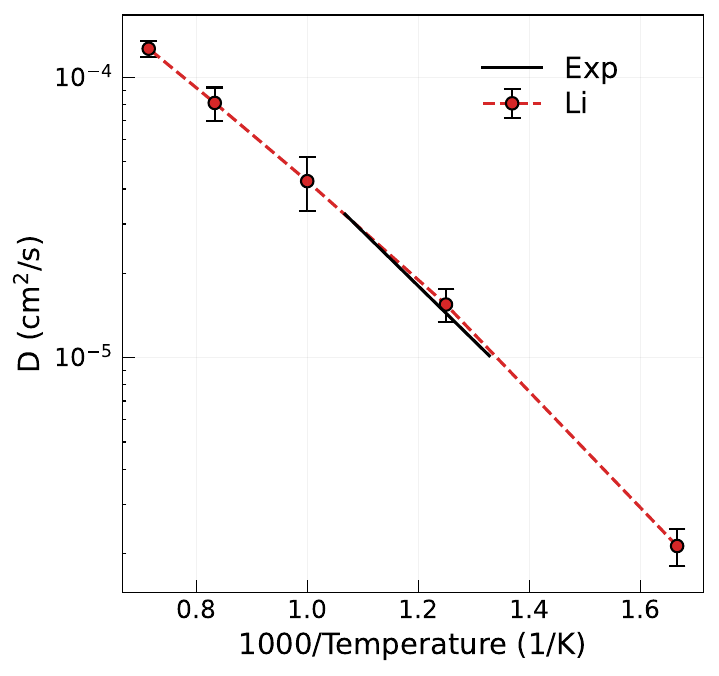}} \\ (a)
	\end{minipage}
 	\begin{minipage}[h]{0.49\linewidth}
		\center{\includegraphics[trim={0cm 0cm 0cm 0cm}, clip, width=0.95\linewidth]{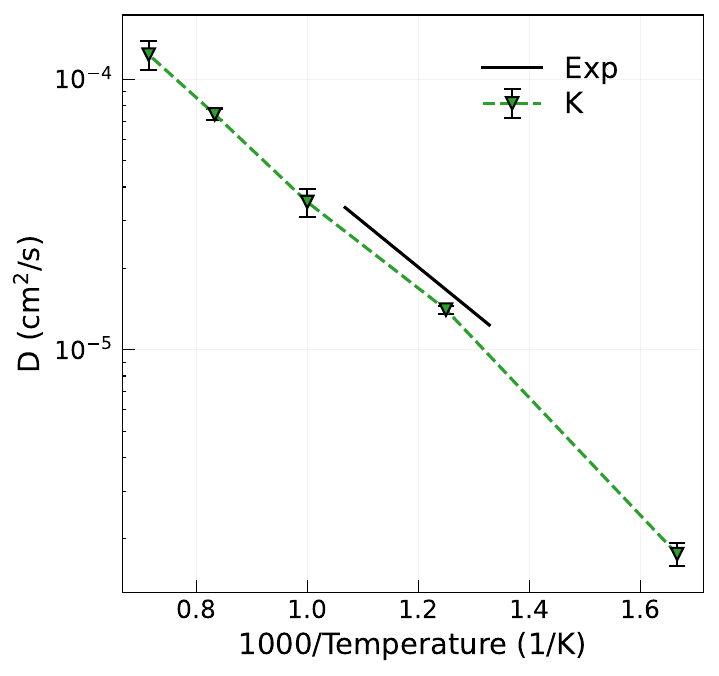}} \\ (b)
	\end{minipage}
  	\begin{minipage}[h]{0.49\linewidth}
		\center{\includegraphics[trim={0cm 0cm 0cm 0cm}, clip, width=0.95\linewidth]{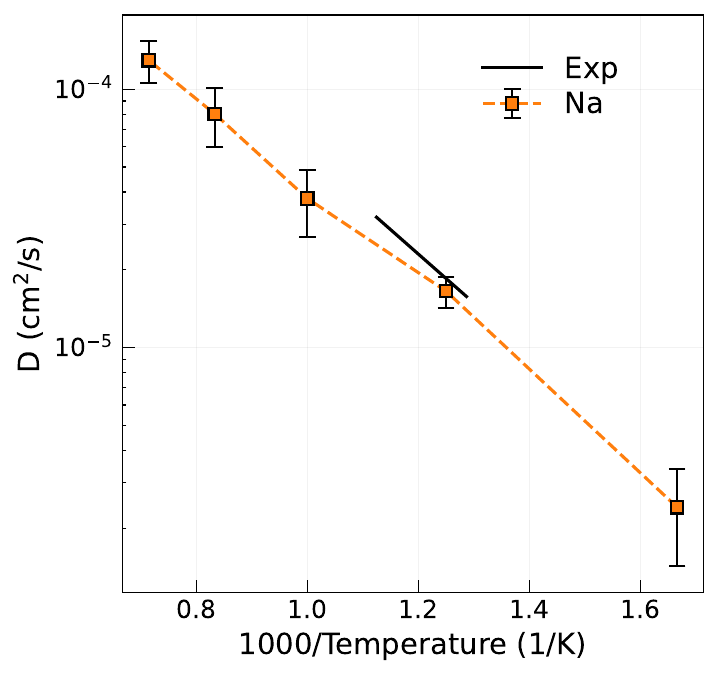}} \\ (c)
	\end{minipage}
  	\begin{minipage}[h]{0.49\linewidth}
		\center{\includegraphics[trim={0cm 0cm 0cm 0cm}, clip, width=0.95\linewidth]{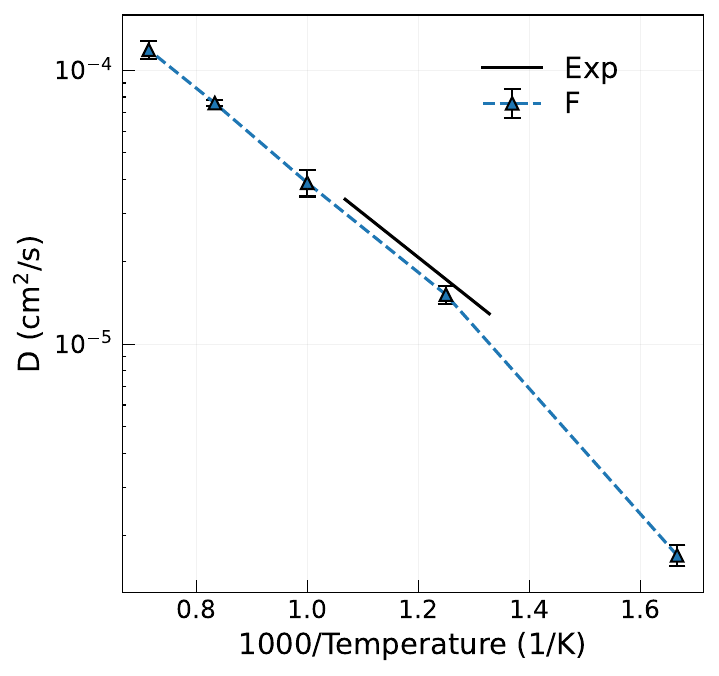}} \\ (d)
	\end{minipage}
	\caption{Temperature dependence of the self-diffusion coefficients for (a)~Li, (b)~K, (c)~Na, (d)~F. Experimental data is taken from~\cite{Umesaki1981}. Error bars are calculated from the standard deviation among 5 separately trained MTPs.}
	\label{fig:diffusivity}
\end{figure*}

Self-diffusion of atoms in a melt is another fundamental property of liquid dynamics that provides important structural information~\cite{Rollet2009}. Fig.~\ref{fig:diffusivity}(a,~b,~c,~d) show the temperature-dependence of self-diffusion coefficients of F, Li, Na, and K on a logarithmic scale, obtained from MTP-MD and experiment~\cite{Umesaki1981}. Although the experimental data is limited to a relatively short temperature intervals, for all four atomic species the calculated values and the slope of temperature dependencies are both close to the experimental results. Such results further validates our approach and demonstrated benefits of using atomistic simulations to accompany the experiment. 

\subsection{Density}

We next study the temperature dependence of the density as one of the most technologically important thermophysical characteristics of the molten salt. While experimental measurements are feasible, computational modeling enables fast and inexpensive exploration of various salt mixtures with diverse components and compositions. The density of molten FLiNaK has been experimentally measured at different temperatures \cite{Romatoski2017, An2017, Powers1963}, and recent efforts have employed ML potentials based on neural networks to calculate it \cite{Lee2021}.

\begin{figure}
	\centering 
	\includegraphics[width=0.45\textwidth]{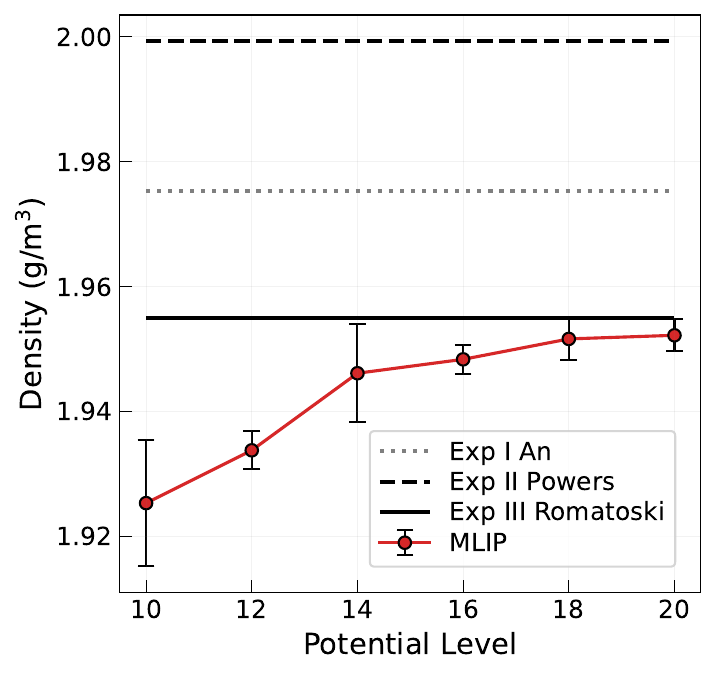}	
	\caption{Density of FLiNaK calculated at T=1000K using MTP with different levels of complexity (number of parameters). The experimental data is taken from~\cite{An2017, Powers1963, Romatoski2017}. Error bars represent one standard deviation calculated over 5 separately trained MTPs.} 
	\label{fig:density_pot_level}
\end{figure}

We perform density calculations of molten FLiNaK, using the same simulation cell as for the diffusion calculations (comprising 3200 atoms at eutectic composition). First, we investigated the influence of different levels of MTP (i.e., varying numbers of parameters) in predicting the density of melt at T=1000K. Fig.~\ref{fig:density_pot_level} illustrates that beyond the level 16, increasing the complexity of MTP does not substantially enhance density values towards experimental results, while computational costs grow. For all subsequent calculations in this study, we employed the potential with level 16. Notably, such a clear convergence of the potential complexity and the property of interest (density in this case) as in Fig.~\ref{fig:density_pot_level} is, however, not always observed. We also note that fitting of the MTPs is based on the Broyden-Fletcher-Goldfarb-Shanno method, and the optimized parameters depended on the initialization of MTP parameters, resulting in varying values and errors for the targeted property. To address this variability, we conducted 5 optimization sessions for each potential level, with error bars depicted in Fig.~\ref{fig:density_pot_level}. Our findings indicate that the level of as low as 10 can yield reasonably accurate results for molten FLiNaK density. The errors on energies and forces exhibit nearly identical patterns across different potential levels (from 10 to 20), aligning with observations in a recent study of molten LiF-BeF$_{2}$, where MTP was also used~\cite{Attarian2022}. 

\begin{figure}
	\centering 
	\includegraphics[width=0.45\textwidth]{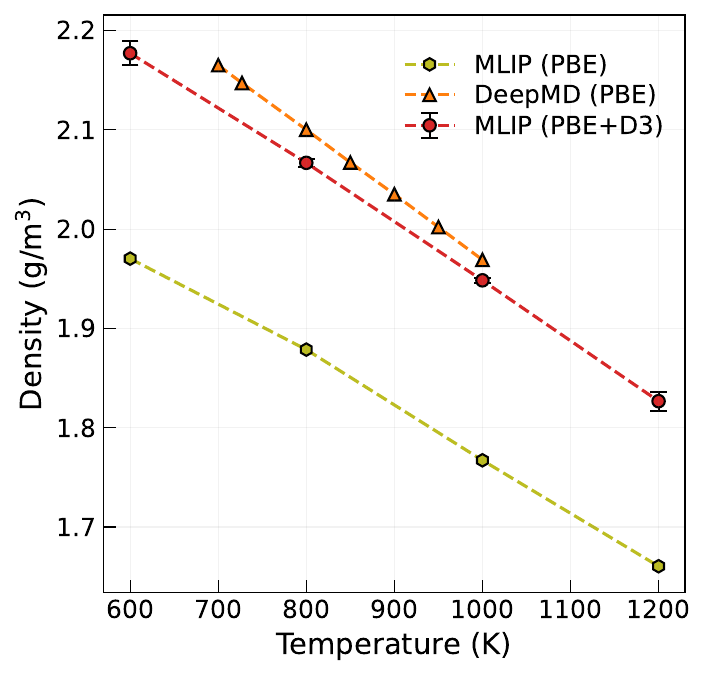}	
	\caption{Temperature dependence of the density of FLiNaK calculated using MLIP. Olive disks represent results of MTP calculated on a dataset in which D3 dispersion corrections was not taken into account, while red disks represent MTP, which was trained on dataset with the D3 correction~\cite{Grimme2010}. Density calculated using DeepMD (orange triangles) is taken from~\cite{Lee2021}. Error bars represent one standard deviation calculated over 5 separately trained MTPs.} 
	\label{fig:density_mlip_theory}
\end{figure}

Next, we proceed toward the comparison of our results with previous theoretical calculations. Recently, DeepMD potential was used to study molten FLiNaK~\cite{Lee2021}. We performed density calculations for temperatures T=600~K, 800~K, 1000~K, 1200~K. Fig.~\ref{fig:density_mlip_theory} reveals an overall agreement between our results and those reported in~\cite{Lee2021}. Notably, the configurations used for potential training in~\cite{Lee2021} were sampled at a fixed experimental density in the NVT ensemble, whereas we performed AIMD simulations with the NPT ensemble.

Given that FLiNaK is an ionic compound, accounting for long-range interactions becomes crucial in atomic dynamics calculations. Although the functional form of MTP generated by the MLIP-2 package currently lacks explicit terms for dispersion corrections, our tests of the convergence of energies and forces with respect to the cutoff radius of interactions showed no substantial improvements beyond a cutoff radius of 5.5~$\AA$. This distance is deemed sufficient to include multiple neighbor shells in the liquid, with negligible electrostatic interactions expected beyond this distance for near-equilibrium FLiNaK configurations. While directly incorporating long-range interactions treatment into potential construction could potentially enhance simulation accuracy, such functionality has not been developed yet.

To the best of our knowledge, the literature lacks an in-depth exploration of the impact of long-range interactions on thermophysical properties calculations of molten salts. A study \cite{Frandsen2020} demonstrated that including vdW-corrections increased the error in density determination by 4\%. In our case, Fig.~\ref{fig:density_mlip_theory} illustrates how a potential trained on data with the DFT-D3 correction leads to slightly different results (about a 10\% difference) compared to a potential trained without the the D3 correction. In our case, including long-range interactions during the dataset generation step improved agreement with experimental results. This test was performed by generating a dataset using the active learning methodology described earlier.

\begin{figure}
	\centering 
	\includegraphics[width=0.45\textwidth]{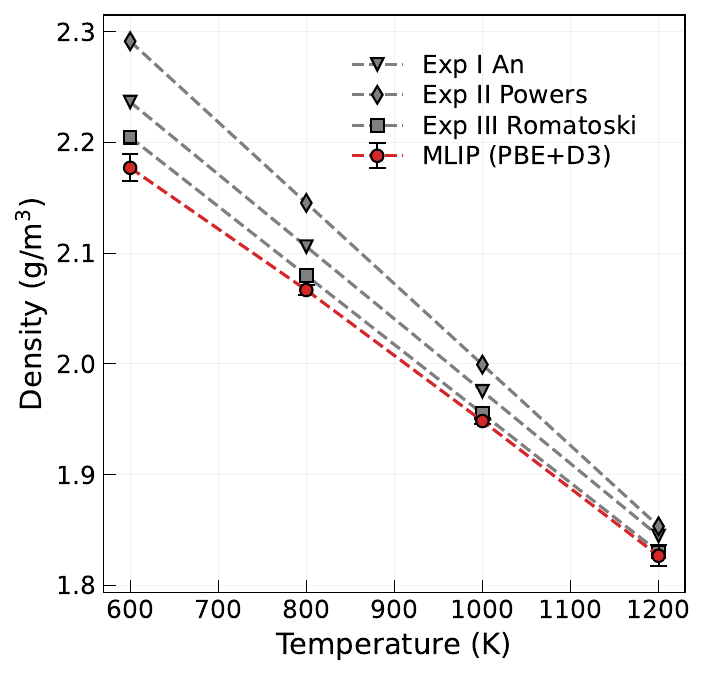}	
	\caption{Temperature dependence of the density of FLiNaK. The experimental data is taken from~\cite{An2017, Powers1963, Romatoski2017}. Error bars represent one standard deviation calculated over 5 separately trained MTPs.} 
	\label{fig:density_mlip_exp}
\end{figure}

Finally, Fig.~\ref{fig:density_mlip_exp} presents a comparison of temperature-dependent density of molten FliNaK obtained using MTP-MD with several experimental results~\cite{An2017, Powers1963, Romatoski2017}. Our findings align well with the experimental results, despite not fixing experimentally-known density during the preparation of the training set. Given that experimental density may not be available for other salts, we emphasize this as an appropriate approach for conducting every MLIP benchmarks.

In the following subsections we will demonstrate how MTP can be used in conjunction with the Green-Kubo method~\cite{Kubo1957, Green2004} to calculate viscosity of molten salt and with the M\"{u}ller-Plathe non-equilibrium method~\cite{Plathe1997} for thermal conductivity calculations. It should be noted that due to the high computational cost of conducting viscosity and thermal conductivity calculations the reported values are presented without error bars.

\subsection{Viscosity}

We calculated viscosity using the Green-Kubo~(GK) approach~\cite{Kubo1957, Green2004}, employing the implementation available in the LAMMPS package. The GK approach involves the calculation of viscosity through the integral of the auto-correlation function of the off-diagonal elements of the stress tensor, expressed by the following relation:
\[
\eta = \frac{V}{k_{B}T} \int_{0}^{\infty} \braket{P_{\alpha\beta}(t)P_{\alpha\beta}(0)}dt,
\]
where $\eta$ is viscosity, $k_{B}$ is the Boltzmann constant, and $P_{\alpha\beta}$ are the off-diagonal elements of the stress tensor. 

In our viscosity calculations we again utilized a simulation cell containing 3200 atoms, with a simulation time step of 1~fs. For all temperatures, an auto-correlation time of 10~ps was selected ensuring the convergence of the auto-correlation function of the diagonal stress components to zero.
After the initial equilibration at each temperature, the simulation was extended for 5~ns, employing the NVE ensemble.

\begin{figure}
	\centering 
	\includegraphics[width=0.48\textwidth]{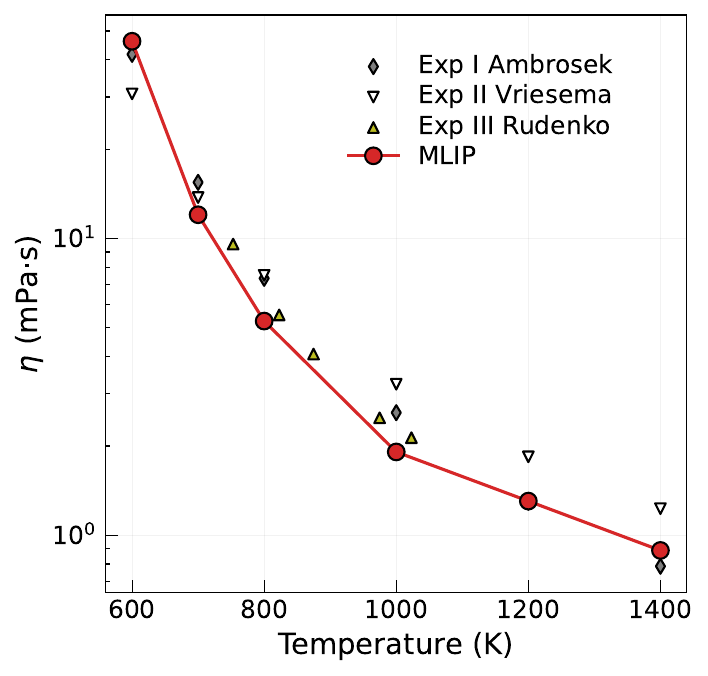}	
	\caption{Temperature dependence of the viscosity of molten FLiNaK (eutectic composition). Experimental data is taken from~\cite{Ambrosek2009, Vriesema1979, Rudenko2022_2}.} 
	\label{fig:viscosity}
\end{figure}

Fig.~\ref{fig:viscosity} illustrates the temperature-dependent viscosity of FLiNaK calculated here and experimentally~\cite{Ambrosek2009, Vriesema1979, Rudenko2022_2}. Our results exhibit excellent agreement with the experimental data. The relationship between viscosity and temperature established in our work is described by the following equation:
\[
    \eta = 0.00543 \cdot \exp\left(\frac{5422}{T} \right) ~ [{\rm mPa} \cdot {\rm s}].
\]

\begin{figure}
	\centering 
	\includegraphics[width=0.5\textwidth]{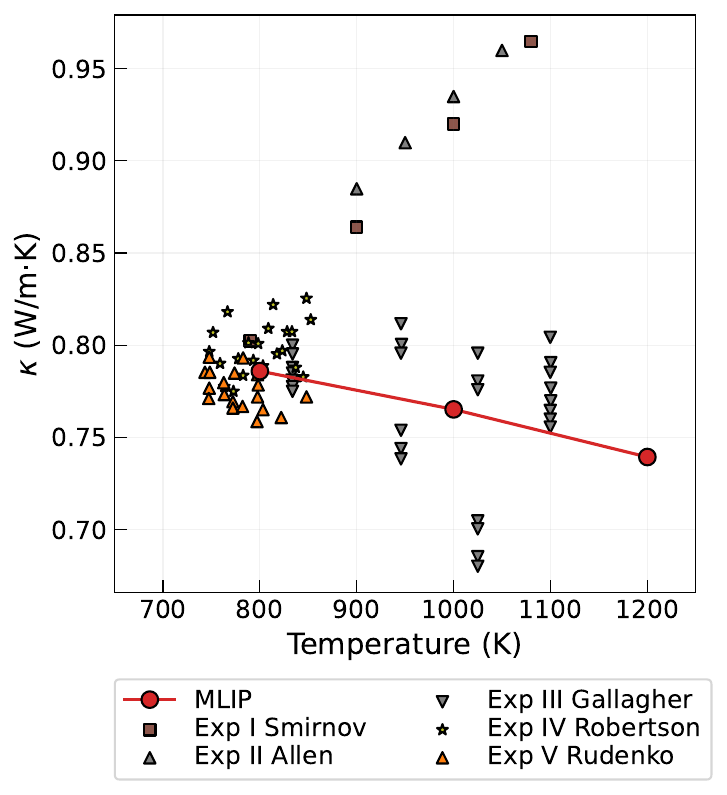}	
	\caption{Temperature dependence of the thermal conductivity of molten FLiNaK (eutectic composition). The experimental results are taken from~\cite{Smirnov1987, Sohal2010, Gallagher2022, Robertson2022, Rudenko2022}. In \cite{Smirnov1987} $\kappa$ was measured with a correction of $\pm$0.012 W/mK, introducing an uncertainty of 1-5\%.} 
	\label{fig:thermal_conductivity}
\end{figure}

\subsection{Thermal Conductivity}

Before delving into the discussion of our heat transport calculations, it is essential to provide an overview of the available experimental data on the temperature dependence of a thermal conductivity (labeled as $\kappa$ from now on). The earliest works provide $\kappa$ values ranging from 1.3~W/mK to 4.5~W/mK,~\cite{Vriesema1979, Grele1953, Hoffman1955}. Some experimental works~\cite{Smirnov1987, Sohal2010} indicate a slight increase in thermal conductivity with temperature, as depicted in Fig.~\ref{fig:thermal_conductivity}. Work~\cite{Williams2006} indicated that the thermal conductivity at 973~K will be in the range 0.6–1.0~W/mK. Most recent experiments~\cite{Gallagher2022, Robertson2022, Rudenko2022} agree in suggesting a slight decrease of $\kappa$ with temperature, indicating that the thermal conductivity value should be below 1~W/mK. 

We determined thermal conductivity of molten FLiNaK using the M\"{u}ller-Plathe non-equilibrium method~\cite{Plathe1997}. It was demonstrated in~\cite{Pan2021} that this approach allows computing heat transport in molten salts in a better agreement with experimental values, comparably to the Green-Kubo method. We used a time step of 0.5~fs and the supercell with 4000 atoms (2.6~nm~$\times$~2.9~nm~$\times$~23~nm dimensionality) and a kinetic energy swap rate of 1 in every 1000~steps. For each temperature, after the initial equilibration in the NPT ensemble for 10~ps, the simulations were done for 4~ns in the NVE ensemble. Fig.~\ref{fig:thermal_conductivity} shows results of our calculations and mentioned above experimental data. The linear fit of our results lead to the following equation for the thermal conductivity:
\[
    \kappa = 0.880 - 0.116 \cdot 10^{-3} T \text{ [W/m} \cdot \text{K]}
\]

As demonstrated in Fig.~\ref{fig:thermal_conductivity}, although the experimental data is slightly scattered, our calculated values agree well with the most recent experimental works~\cite{Gallagher2022, Robertson2022, Rudenko2022}. Taken into account the magnitude of thermal conductivity changes with temperature, we are of the opinion that it is reasonable to state that the thermal conductivity in FLiNaK stays nearly constant with temperature grows and its value is below 1~W/mK. 

\section{Summary and conclusions}

In this work, we utilized a MTP and assessed its performance for the calculations of thermophysical properties of molten FLiNaK in a range of temperatures. Employing an active learning approach enabled the rapid training of a robust potential capable of predicting density, diffusion, viscosity, and thermal conductivity with near \textit{ab initio} accuracy. While previous computational works demonstrated good agreement with experimental data, the MTP, coupled with active learning, stands out for its ability to deliver results much faster, thanks to its efficient utilization of data. This significantly reduces the computational cost compared to previous studies employing alternative MLIPs. This breakthrough paves the way for swift yet precise exploration of thermophysical properties across various salt systems.

Our analysis, encompassing radial distribution function and diffusivity calculations, demonstrates the MTP's accuracy in predicting local structures of molten FLiNaK. The extrapolation of the potential to larger systems, achieved using a smaller simulation cell during the ML potential training, underscores its versatility in handling diverse local configurations. While the MTP has a medium-range cutoff distance (up to 5.5~$\AA$ in this study) and does not explicitly consider long-range interactions, our results indicate that it performs well in ionic systems. The absence of explicit charge treatment in the MTP, justified by local ionic solvation shells limiting long-range interactions, makes it a useful tool for fast and accurate computations of molten salt's thermophysical properties. However, certain conditions must be met for optimal results. Our findings suggest that accounting for van der Waals dispersion during the generation of training data improves predictions of the temperature dependence of density. While the potential currently lacks an implicit treatment of long-range interactions, recent research~\cite{Ko2021, Gao2022} highlights the possible benefits of their explicit inclusion in charge systems. Such features should be included in the next generation of MTPs to improve modeling accuracy.

We also note that utilization of different DFT approximations might improve the results for molten salts calculations with MLIPs. In particular, it was recently shown in~\cite{Tisi2021} that training of a MLIP on the dataset generated at the DFT-PBE level of accuracy leads to the predicted values of the thermal conductivity of water 60\% larger than in experiments. Retraining on the dataset generated with the strongly constrained and appropriately normed semilocal density functional (SCAN)~\cite{Sun2015, Sun2016} decreases errors by two times. Proper benchmarking of the exchange-correlation density functional influence on the evaluated properties of molten salts is, to the best of our knowledge, currently missing in literature and have to be explored in future works.

\section*{Acknowledgements}
N.R. and A.S. acknowledge funding from the Russian Science Foundation (Project No. 23-13-00332).

\appendix

\bibliographystyle{elsarticle-harv} 
\clearpage
\bibliography{example}






\end{document}